# Quantum Reality, Complex Numbers and the Meteorological Butterfly Effect

by

T.N.Palmer

European Centre for Medium-Range Weather Forecasts
Shinfield Park, RG2 9AX, Reading
UK

tim.palmer@ecmwf.int

**Abstract**

Meteorology is a wonderfully interdisciplinary subject. But can nonlinear thinking about predictability of weather and climate contribute usefully to issues in fundamental physics? Although this might seem extremely unlikely at first sight, an attempt is made to answer the question positively. The long-standing conceptual problems of quantum theory are outlined, focussing on indeterminacy[1] and non-local causality; problems that led Einstein to reject quantum mechanics as a fundamental theory of physics. These conceptual problems are considered in the light of both low-order chaos and the more radical (and less well known) paradigm of the finite-time predictability horizon associated with the self-similar upscale cascade of uncertainty in a turbulent fluid. The analysis of these dynamical systems calls into doubt one of the key pieces of logic used in quantum non-locality theorems: that of counterfactual reasoning. By considering an idealisation of the upscale cascade (which provides a novel representation of complex numbers and quaternions), a case is made for reinterpreting the quantum wave-function as a set of intricately-encoded binary sequences. In this reinterpretation, it is argued that the quantum world has no need for dice-playing deities, undead cats, multiple universes, or "spooky action at a distance".

**1. Introduction**

Weather and climate affect the lives of virtually everyone on the planet. It is not surprising, therefore, how interdisciplinary is the science of meteorology, with clear quantitative links to many applied sciences (eg Palmer et al, 2004). But is it possible that interdisciplinary links might also exist "back" towards more fundamental physics? More specifically, could nonlinear thinking about predictability of weather help reformulate quantum theory in such a way as to help solve the conceptual and foundational difficulties which made the theory so difficult for Einstein to accept, and which continue to plague the theory to this day (Penrose, 2004)? On the face of it, this seems an utterly preposterous idea - what could the largely classical and familiar world of meteorology have to say about counter-intuitive notions like wave-particle duality, quantum non-locality, parallel universes and the like?

Notwithstanding such entirely reasonable pre-conceptions, the purpose of this article is to suggest that application of nonlinear meteorological thinking may indeed provide fresh insights on the foundational problems of quantum theory, as summarised in section 2. On this, the 100[th] anniversary of the publication of Einstein's seminal work on quantum theory and the photoelectric effect, we will focus on the two main concerns which ultimately led him to reject this theory: indeterminacy ("I cannot believe that God plays dice with the cosmos") and, more importantly, the notion of non-local causality, which Einstein referred to as "spooky action at a distance".

In section 3 the validity of a key assumption required to demonstrate non-local causality in quantum theory is called into doubt from the perspective of a toy universe governed by the prototypical Lorenz (1963) model of low-order chaos. On the other hand, it is also shown in section 3 that low-order chaos cannot itself provide a solution to these quantum foundational problems. In section 4 is discussed the apparent finite-time predictability horizon associated with 3D inviscid fluid motion, referred to as the "meteorological butterfly effect". The existence of this horizon is associated with a self-similar upscale cascade of uncertainty and is quite different from that associated with low-order chaos. In section 5, an idealised representation of the upscale cascade is formulated using permutation operators that have the same multiplicative properties as the unit complex numbers. Complex numbers play an essential role in describing the evolution of the quantum wave-function, and in section 6, this reinterpretation of complex numbers leads to a reformulation of the quantum wave-function as a set of intricately-encoded binary sequences ("quantum DNA"). This reinterpretation allows the arguments in section 3 to be used to reject both quantum indeterminacy and quantum non-local causality.

---

[1] A glossary of some of key terms used in the paper is given in the appendix

The technical work on which this paper is based has been published (Palmer, 2004) in a journal not widely read in meteorological circles. However, as the ideas put forward were motivated by idealised problems of predictability in meteorology, they might be of interest to philosophically-minded members of the American Meteorological Society.

This paper assumes no prior knowledge of quantum theory. A glossary of the key terms used in the paper is given in the appendix.

## 2. Some Quantum Background

Quantum theory is the most successfully tested, yet least-well understood of all physical theories (see for example, Penrose, 2004). Einstein's dissatisfaction with quantum theory is well known; this section briefly summarises the two key reasons for such dissatisfaction: indeterminacy and non-local causality

According to quantum theory, a quantum system (eg a photon) is described by a quantity called the wave-function $|\psi\rangle$. When the quantum system is not being "observed" eg interrogated by some laboratory apparatus, $|\psi\rangle$'s evolution in time is described by a deterministic linear differential equation

$$i\hbar \frac{\partial |\psi\rangle}{\partial t} = H|\psi\rangle \qquad (1)$$

known as the Schrödinger equation. Here $\hbar$ is a form of Planck's constant and $H$ is a linear operator known as the Hamiltonian, whose classical (ie non-quantum) form is well-known to mathematical meteorologists as an expression for total energy. For the purposes of this paper, an absolutely central point about the Schrödinger equation is that it explicitly contains reference to the complex number $i = \sqrt{-1}$.

On the other hand, when we try to use quantum theory to predict the outcome of some possible measurement (eg to determine which path a photon takes through an interferometer, see below) quantum theory says that $|\psi\rangle$ evolves non-deterministically; all that can be predicted is the probability of one of a number of possible outcomes. Stochastic generalisations of the Schrödinger equation can be formulated to account for this indeterminacy (Percival, 1998). In this sense, the indeterminism of quantum theory appears to arise from some external interaction of $|\psi\rangle$ with the classical world (which may or may not include laboratory apparatuses). However, notwithstanding the philosophical difficulties associated with the notion of randomness (Stewart, 2004), surely it is unsatisfactory for a fundamental theory of physics to have to assume, *ab initio*, such a pre-existing classical world? This problem is brought into focus in cosmology. Cosmologists who wish to investigate $|\psi\rangle$ for the whole universe do not have the luxury of invoking an external classical world.

Because of this, some prominent quantum theorists believe that these probabilities arise because the universe splits into multiple universes every time a quantum interaction occurs (Deutsch, 1988). For example, the number of universes where outcome O is observed, is in proportion to the probability of occurrence of O, as given the "normal rules" of quantum theory. Many people find this so-called "many-worlds" interpretation too bizarre to accept, but equally hard to reject objectively.

One of the most famous experiments which illustrates quantum strangeness is Young's two-slit experiment (Fig 1a). This clearly demonstrates the wave nature of light, diffraction at the two slits causing an interference pattern on the back screen as the coherent beams of light emanating from the two slits combine. If the intensity of the light source is reduced so that the source only emits one photon at a time, a photon is never observed to split in two, ie travel through the two slits at the same time; rather, individual photons are observed to go through one slit or the other. Despite this, when no attempt is made to find out which slit a given photon goes through, photons are never observed at a minimum of the classical interference pattern. The mystery is this: how do these single photons "know" never to travel to a position of destructive interference?

Standard quantum theory accounts for this mathematically by making use of the linearity of the Schrödinger equation. If $|\psi_{top}\rangle$: "photon travels through top slit" and $|\psi_{bottom}\rangle$:"photon travels through bottom slit" are solutions of the Schrödinger equation, then so are complex linear superpositions $(|\psi_{top}\rangle + e^{i\lambda}|\psi_{bottom}\rangle)/\sqrt{2}$. If we try to observe the existence of a photon near one of the slits, standard quantum theory says that this superposed wave-function reduces indeterminately to either $|\psi_{top}\rangle$ or $|\psi_{bottom}\rangle$ with equal probability. However, if we do not try to observe the photons travelling through the slits, then the superposed states remains a coherent entity until the photon (or lack of it) is observed at the back screen. The interference pattern at the back screen arises from variations in the complex phase factor $\lambda$.

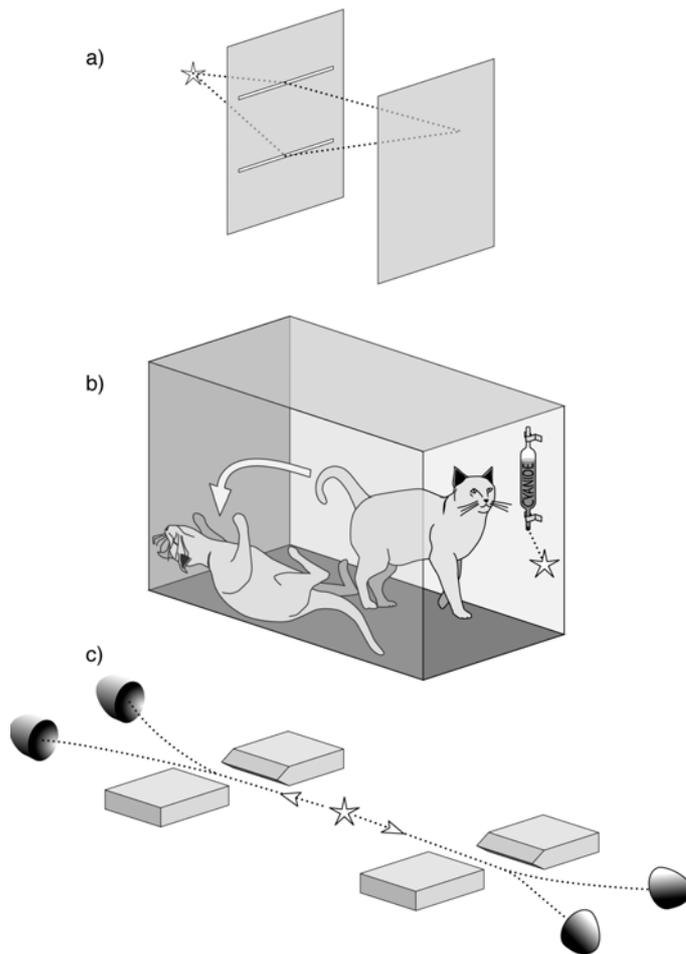

**Figure 1**. *Some conceptual problems in quantum theory. a) Young's two-slit experiment when the intensity of the light source is so low that only one photon is emitted at a time. Since the photons are never observed to split into two, how does any one photon know never to travel to a minimum of the interference pattern? b) Schrödinger's cat in a closed box containing a phial of cyanide. If a radioactive atom decays within a certain period, the phial breaks and kills the cat. According to quantum theory, within this period of time, the cat is in a linearly superposed state of aliveness and deadness. c) A spin-0 source emits two "entangled" spin-1/2 particles in a suitably superposed spin state. According to quantum theory, measuring the spin of the left-hand particle instantaneously causes the spin state of the right-hand particle to a definite "up" or "down", no matter how far apart the two measuring devices are, contrary to the spirit of Einstein's theory of relativity.*

Schrödinger himself realised the ludicrousness of the notion of "superposition". To illustrate this, he considered a cat in a closed box containing a phial of deadly cyanide gas (Fig 1b). The phial breaks if a radio-active atom decays within a period of time. According to quantum theory, the state of the atom within this period is again in a linearly superposed state (of decay and non-decay). On this basis, the cat too is somehow in a linear superposition of aliveness and deadness! Only by us observers opening the box and looking inside, does the cat non-deterministically evolve (by a supposed shake of God's dice) from this "undead" state, to one of definite aliveness or deadness. If we are to believe the standard rules of quantum theory, it is our curiosity that kills the cat! Why this should be, remains mysterious. Why we do not observe superposed states remains a very controversial topic, even amongst quantum experts.

Putting these two problems together creates the apparent effect of non-local causality: what Einstein referred to as "spooky action at a distance". This was his most profound objection to quantum theory. Two spin-1/2 particles (eg electrons) are emitted from a zero angular-momentum source in a type of correlated superposed state known as an "entangled" state (Fig 1c). Let's say that the spin of the left-hand particle is then measured to be "up", relative to some direction **n**. According to standard quantum theory, this instantaneously causes the entangled spin state to reduce to something definite, so that the spin of the distant particle is necessarily "down" with respect to **n**. The phrase "spooky action at a distance" refers to the notion that a spin measurement on the left-hand particle has apparently instantaneously caused the spin of the right-hand particle to change from having an indeterminate value, to having some definite value, no matter how far apart these two particles are.

No wonder Einstein was upset with this notion; invoking instantaneous action at a distance undermines his most valuable contribution to physics: the theory of relativity (which says, broadly speaking, that the effects of some localised cause can propagate no faster than the speed of light). Einstein believed that there must be some underlying theory, deeper than conventional quantum theory, which was both realistic and local; realistic in the sense that quantum states can always be assigned definite values (ie up or down and not some strange combination of both), and local in the sense that distant measurements cannot instantaneously cause these spin values to change, ie no "spooky action at a distance". However, a celebrated mathematical result known as Bell's theorem (Bell, 1993) is usually interpreted as saying that if Einstein was right, the correlations between spin measurements of such entangled particle pairs must satisfy a certain inequality. Experimentally, this inequality is known to be violated. Hence, conventional wisdom has it that Einstein was wrong not to believe in spooky actions at a distance. (A readable version of Bell's theorem for the non-specialist can be found in Rae; 1986)

Whilst Bell's theorem appears to imply some form of non-local causality in quantum theory, there is something profoundly slippery going on here, because it can also be shown that quantum systems cannot be used to send information faster than the speed of light. What on earth is going on?

**3. Counterfactual Definiteness and the Prototype Model of Weather Chaos**

In this section the prototype model of weather chaos is used to call into question an implicit assumption in the proof of Bell's theorem - one rarely mentioned explicitly in text-book proofs. This assumption is rather metaphysical (consistent with the slipperiness mentioned at the end of the previous section) and is called "counterfactual definiteness". The notion of "counterfactual definiteness" is used to define what is meant by non-local causality: that some remote "cause" can lead to an instantaneous "effect" locally. By questioning the notion of counterfactual definiteness, we are in turn calling into doubt the meaningfulness of the notion of non-local causality.

Fig 1c showed the situation when both the left and right hand particles of an entangled particle pair were measured with magnets oriented in the same direction **n**. However, in order to establish Bell's theorem, we need to consider correlations between pairs of measurements when the magnets have different orientations, let's say **n** for the left-hand magnets and **n'** for the right-hand magnets. It is also necessary to assume that it is meaningful to ask: what would the spin of a left-hand particle have been had we actually measured it with magnets oriented in the **n'** direction (or, conversely, what would the spin of the right-hand particle have been had we actually measured it with magnets oriented in the **n** direction)?

Note that by definition this question could never be actually answered experimentally. In fact it is an example of a counterfactual question, a question about things that didn't happen, but our intuition suggests might have happened. It is easy to think of other examples of counterfactual questions. Would the world have gone to war in the last century if the assassin's bullet had missed Archduke Franz Ferdinand in Sarajevo in 1914? Would the weather in London have been sunny today if some Amazonian butterfly had not flapped its wings exactly one month earlier? Although we can't know the answer to these questions for sure, we nevertheless feel intuitively that each should in principle have answers. But is intuition a reasonable guide in these matters?

If we analyse the operational meaning of such counterfactual questions, we are required to imagine, at some instant in time, a hypothetical dynamically-unconstrained perturbation to the actual universe, affecting only one small part of the universe, keeping the rest unchanged. Hence we must imagine a localised perturbation to the path of a specific speeding bullet in Sarajevo, or to the wings of a specific butterfly in Amazonia. In the case of Bell's theorem, the corresponding imagined perturbation is one that changes the orientation of the magnets, keeping unchanged the rest of the universe, including in particular the particle whose spin is being measured. The counterfactual questions are then in principle answerable by somehow integrating forward in time the dynamical equations of motion of the universe from these hypothetical perturbed states.

Now if we were imagining gigantic hypothetical perturbations, such as would move whole galaxies around willy-nilly, we might well worry whether our imagination was being consistent with the laws of physics! However, issues of physical consistency seem less important for the counterfactuals above, because the imagined perturbations can be considered in some sense to be arbitrarily small. (For example, if you think the flap of a butterfly's wing is too large a perturbation, imagine instead some even smaller perturbation to a neuron in the butterfly's brain that somehow causes it to not flap at the key moment). Chaos theory suggests that, in this respect, our intuition may indeed be an unreliable guide.

Consider, for example, the prototypical Lorenz (1963) equations

$$\dot{X} = -\sigma X + Y$$
$$\dot{Y} = -XZ + rX - Y \qquad (2)$$
$$\dot{Z} = XY - bZ$$

whose attractor is illustrated (schematically) in Figure 2. (Here, *X,Y,Z* are the components of the state vector of the Lorenz model. The parameters *r*, *b* and σ are considered fixed, and have values 28, 8/3 and 10 respectively in Figure 2.)

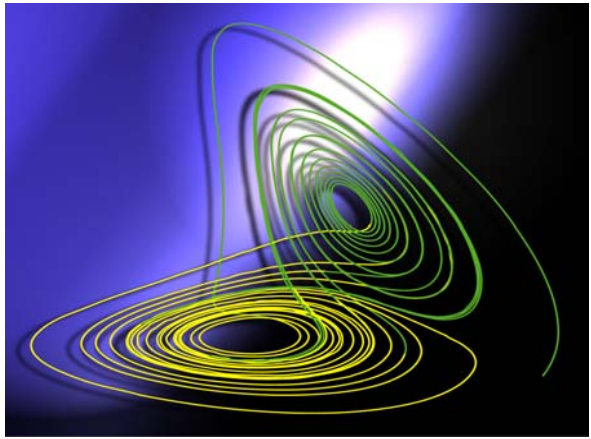

*Figure 2. An illustration of the Lorenz (1963) attractor.*

Following this, consider a "Lorenzian universe" whose two "laws of physics" are i) equation (2) and ii) a "cosmic" initial state lying of the attractor of (2). Now, at some time *t*, imagine some hypothetical dynamically-unconstrained perturbation which changes one component of the Lorenzian state vector, keeping the other components fixed. Specifically, let $X(t) \rightarrow X(t)+\delta X$ keeping $Y(t)$ and $Z(t)$ unchanged. Similar to the counterfactual perturbations above, $\delta X$ has been posited without regard to the "laws of physics" of our Lorenzian universe. Now, because of the fractal nature of the attractor, then no matter how small is $\delta X$, the resulting perturbed state is almost certainly (ie with probability one with respect to the continuum measure of phase space) off the attractor. Hence the imagined perturbed state $(X(t)+\delta X, Y(t), Z(t))$ is inconsistent with the second Lorenzian "law of physics". That is to say, if were to ask the counterfactual question: would *Y* have been positive at $t_2 > t_1$ if X had differed from its actual value at $t_1$ by some imagined tiny amount $\delta X$, then, according to the laws of physics of the Lorenzian universe, this counterfactual question is almost certainly neither true nor false.

This argument calls into doubt counterfactual reasoning, and hence the notion of causality based on counterfactual reasoning. As discussed below, without counterfactual definiteness we cannot use the experimental violation of the Bell inequalities to infer non-local causality, and hence "spooky action at a distance".

However, is there any evidence that low-order chaos somehow underlies the fundamentals of quantum theory? This question has been discussed by the experts. For example, David Deutsch, one of the founding fathers of the quantum computer (which one day, who knows, may be used to make weather forecasts) explains in his popular book "The Fabric of Reality"(Deutsch, 1998):

"Chaos theory is about limitations on predictability in classical physics, stemming from the fact that almost all classical systems are inherently unstable…It is about an extreme sensitivity to initial conditions. …Thus it is said that in principle the flap of a butterfly's wing in one hemisphere of the planet could cause a hurricane in the other hemisphere. The infeasibility of weather forecasting and the like is then attributed to the impossibility of accounting for every butterfly on the planet.

However, real hurricanes and real butterflies obey quantum theory, not classical mechanics. The instability that would rapidly amplify slight mis-specifications of an initial classical state is simply not a feature of quantum-mechanical systems. In quantum mechanics, small deviations from a specified initial state tend to cause only small deviations from the predicted final state. Instead, accurate prediction is made difficult by quite a different effect."

In the above paragraph, Deutsch (who believes strongly in the many-worlds interpretation) is essentially remarking that the Schrödinger equation is linear and stable, whilst chaos is nonlinear and unstable. We appear to have a fundamental incompatibility.

Roger Penrose is one of the few of the leading scientists working in the field who follows Einstein in believing that standard quantum theory is incomplete. He maintains that a satisfactory quantum-compatible theory of gravity will never emerge from the normal rules of quantum theory (a view which would rule out string theory as the basis of a theory of everything). Moreover, Penrose also presents evidence that any underlying theory of the quantum world should be based on some notion of non-computability (Penrose, 1994). A class of (mathematically well-posed) propositions is referred to as non-computable if there is no algorithm that can be guaranteed to determine the truth or falsehood of each member of the class. As an example of non-computability, consider the famous Mandelbrot set (see Fig 2). The point $c$ on the Argand plane is said to belong to the Mandelbrot set if, starting from $z=0$, the sequence of iterates of $|z|$ under $z \rightarrow z^2+c$, does not diverge to infinity. The notion of non-computability arises here because (Blum et al, 1998) no algorithm can decide whether any given point on the plane lies in the Mandelbrot set.

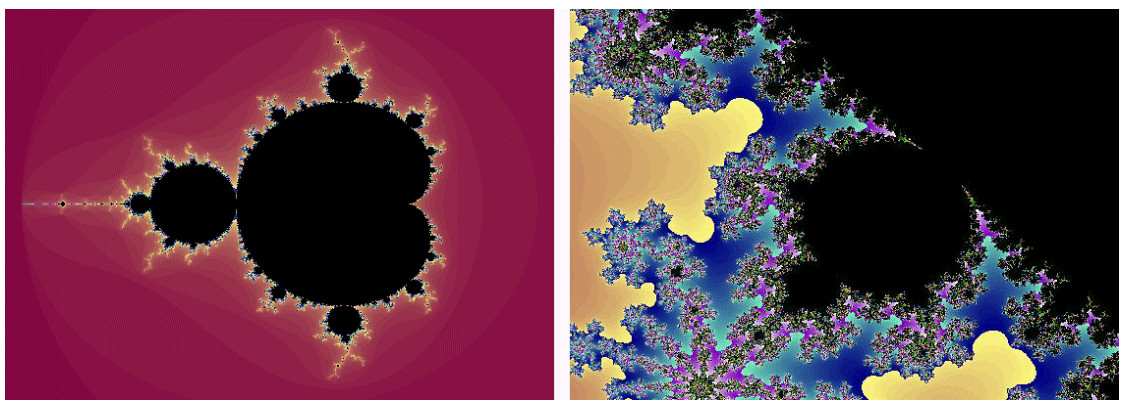

**Figure 3**. *The fractal Mandelbrot set. No algorithm that can decide whether any given point in the plane belongs to the Mandelbrot set (Blum et al 1998). As such the Mandelbrot set is non-recursive, or non-computable. Penrose (1994) has argued that any realistic theory which underpins standard quantum theory must incorporate elements of non-computability.*

If one applied our discussion about the "Lorenz universe", then one could say that it would be impossible to decide algorithmically whether points in the Mandelbrot set continue to so belong under dynamically-unconstrained perturbations. In this sense, non-computability seems also to illustrate the problems with counterfactual reasoning.

However, Penrose (1994) asks:

"Does the phenomenon of chaos provide the needed non-computable physical basis….? An example that is often quoted in this connection is the detailed long-range prediction of the weather."

but then answers:

"I should make it clear that despite such profound difficulties for deterministic prediction, all the normal systems that are referred to as chaotic are to be included in what I call computational…The predicted weather may well not be the weather that actually occurs, but it is perfectly plausible as a weather! .. "

Deutsch and Penrose lie at different ends of the spectrum of informed opinion concerning the meaning of quantum theory. And yet both reject low-order chaos theory as a route to uncovering this "meaning". One problem is that true fractality is not a finite-time property of a chaotic system; you would have to wait an infinite time before the true fractal nature of these dynamics is manifest. The real world of physics cannot wait that long!

**4. The Meteorological Butterfly Effect and the Upscale Cascade**

Are there deterministic systems, studied in meteorology, where total loss of predictability occurs in finite time? Yes! In this section is discussed the notion of the upscale cascade of uncertainty in self-similar multi-scale systems, leading to a much more radical loss of predictability than can be found in low-order chaos.

The term "butterfly effect" is attributed to Ed Lorenz, the father of modern chaos theory (see Lorenz, 1993). However, what I describe as the "meteorological butterfly effect" is not discussed in Lorenz (1963), but in his paper on upscale propagation of small-scale uncertainty in infinite dimensional (eg turbulent) systems (Lorenz 1969). There is an important conceptual difference. As Deutsch correctly points out, at the heart of chaos lies the notion of linear instability associated with exponential growth in amplitudes. By contrast, the meteorological butterfly effect is concerned with the growth in scale of arbitrarily small-scale initial perturbations. Uncertainty in the flap of a butterfly's wings leads to uncertainty in some gust, which leads to uncertainty in some cumulus cloud, which leads to uncertainty in some cyclone. This transfer of uncertainty is profoundly nonlinear; on the scale of a butterfly, a flap of its wings is not necessarily a small-amplitude perturbation.

The arguments in section 3, questioning the notion of counterfactual reasoning, are brought into yet sharper focus for multi-scale systems with low-dimensional attractors and large (potentially infinite) numbers of degrees of freedom. Hence, as above, we can argue that it is virtually certain that a hypothetical dynamically-unconstrained perturbation to a small-scale variable (a flap of a butterfly's wings) leaving all other larger-scale variables fixed, would take the system off the attractor.

However, there is a second reason for considering such multi-scale systems: uncertainty due to a finite-amplitude but arbitrarily small-scale perturbation can propagate nonlinearly upscale, and infect the uncertainty of some given large-scale component of the flow, *in finite time*.

This paradigm appears to apply to 3D fluid turbulence (eg Vallis, 1985). A scaling argument (which the uninterested reader can skip) goes as follows. In the so-called Kolmogorov inertial sub-range, fluid kinetic energy *E(k)* per horizontal wavenumber $k$ scales as $E(k) \sim k^{-5/3}$ so that fluid velocity $u(k) \sim k^{-1/3}$. Let us assume that the time it takes for error at wave number *2k* to infect wave number k (ie to propagate upscale one "octave") is proportional to the "eddy turn over time" $\tau(k) \sim (ku(k))^{-1} \sim k^{-2/3}$. Then the time $\Omega(k_N)$ it takes error to propagate $N_0$ octaves from wavenumber $k_N$ to some large scale $k_L$ of interest is given by the geometric series:

$$\Omega(k_N) = \sum_{n=0}^{N_0-1} \tau(2^n k_L) \qquad (3)$$

Since $\tau \sim 2^{-2n/3} < 1$, $\Omega(k_N)$ tends to a finite limit as $k_N \to \infty$

Mathematically, this can also be viewed as a statement about possible non-uniqueness of solutions of the 3D fluid equations (but not 2D or quasi-geostrophic equations; Vallis 1985): two initial states, which are identical except for some arbitrarily small-scale differences (ie identical in some suitable functional-analytic sense), can diverge finitely on finite scales and in finite time. For the inviscid (yet deterministic) Euler equations of fluid mechanics, there are clear examples of such non-uniqueness (Shnirelman, 1997), but no generic proof exists. Oddly, no counterproof exists for the viscous Navier-Stokes equations. If you can prove or disprove the existence or otherwise of a finite-time predictability horizon for the latter equations, submit your solution to the Clay Mathematics Institute Millenium Prize committee, and win yourself a million dollars (http://www.claymath.org/millennium). Here is a problem on predictability of weather which ranks on a par (financially at least) with the most famous unsolved problems in mathematics, such as the Riemann Hypothesis!

There is more the flavour of Penrose's non-computability in this issue of non-uniqueness, than with ordinary low-order chaos. For example, based on the scaling argument associated with equation (3), forecasting even the sign of the relative vorticity of some large-scale circulation element could not, even in principle, be computed beyond the finite-time predictability horizon. However, to make progress in this direction, we would still have to overcome the apparent incongruity (as highlighted in Deutsch's quote in the section 3) between the complex linear dynamics of the Schrödinger equation, and the real nonlinear dynamics which describes loss of predictability in meteorology.

## 5. The Upscale Cascade and a New Perspective on Complex Numbers

In this section we try to reconcile the apparent incongruity between the complex linear dynamics of the Schrödinger equation, and the real nonlinear dynamics of the upscale cascade. A clue to a possible reconciliation lies in the power-law properties (eg $E(k) \sim k^{-5/3}$) of the upscale cascade, indicative of self-similarity. We see self-similarity quite clearly when we zoom into the non-computable Mandelbrot set, obtained by a simple recursive mapping on the complex plane. As mentioned in section two, complex numbers play an essential role in quantum theory.

Complex analysis is used in many meteorological studies (eg Charney, 1947). On the other hand, I recall the comment of a family friend who told me once that she gave up studying mathematics at school when $\sqrt{-1}$ was introduced. She felt that this was the last straw - corrupting schoolchildren's minds with such nonsensical ideas!

We professionals have a tendency to scoff at such naivety! Given the mathematical consistency (and indeed beauty and power) of the algebra of complex numbers, who cares what $\sqrt{-1}$ "really" means - other than as a symbol with certain properties? And in any case, if we meteorologists really have a conceptual hang-up with complex numbers, we can always solve meteorological problems like the Charney problem (with less efficiency, perhaps) using real analysis.

By contrast, complex numbers lie at the heart of quantum theory; $\sqrt{-1}$ appears explicitly in Schrödinger's equation (1) and related to this, $|\psi\rangle$ is technically an element of a complex Hilbert space . As discussed, the phenomenon of quantum interference is linked to the wavefunction as a complex state vector. Quantum theorists can't be as complacent as meteorologists in ignoring the "meaning" of $\sqrt{-1}$. And yet they are! Having accepted the idea that Schrödinger's cat might exist in a linear superposition of states, I have yet to meet a quantum expert who expresses any further concern that this superposition might indeed be complex! But what is the physical reality of $|\text{Alive cat}\rangle + \sqrt{-1}|\text{Dead cat}\rangle$? My family friend (not to mention the poor cat) would be in a state of apoplexy at the very thought!

Is it possible that there might be some connection between the self-similar upscale cascade and complex numbers? In the last section, we considered the representation of some fluid state by a sequence

$$S = \{a_1, a_2, a_3, a_4, a_5, a_6, a_7, a_8, ..., a_{2^N}\} \tag{5}$$

of coefficients eg of a spherical harmonic basis. In the next section I want to consider representing the wavefunction $|\psi\rangle$ of in terms of sets of sequences such as (5), but where the elements of $S$ are just "1"s or "-1"s. ie $a_j \in \{1, -1\}$. To fix ideas, imagine $S$ to be associated with the first $2^N$ binary digits of a number like $\pi$ or $\sqrt{2}$.

The plan in this section is to reinterpret complex numbers as operators on $S$. To do this, let us first define the negation of $S$ as

$$-S = \{-a_1, -a_2, -a_3, -a_4, -a_5, -a_6, -a_7, -a_8, ...\} \quad (6)$$

so that $-(-S) = S$. Using this, we define the operator *i* to act identically on successive pairs of elements of *S*, negating every second element, and then reversing the order of the elements, so that

$$i(S) = \{-a_2, a_1, -a_4, a_3, -a_6, a_5, -a_8, a_7 ...\} \quad (7)$$

Then it is easily shown

$$i(i(S)) = -S \quad (8)$$

so that *i* can be interpreted as a "square root of minus one". Moreover, by putting

$$i^{1/2}(S) = \{-a_4, a_3, a_1, a_2, -a_8, a_7, a_5, a_6 ...\}$$
$$i^{1/4}(S) = \{-a_8, a_7, a_5, a_6, a_1, a_2, a_3, a_4 ...\} \quad (9)$$

(where $i^{1/2}$ operates identically on successive quadruplets of elements of *S*, $i^{1/4}$ on successive octuplets of elements), the reader can easily verify that

$$i^{1/2}(i^{1/2}(S)) = i(S)$$
$$i^{1/4}(i^{1/4}(S)) = i^{1/2}(S) \quad (10)$$

consistent with multiplication of complex numbers. The pattern of permutations associated with these operators is shown in Fig 4; it is indeed an idealisation of the self-similar upscale cascade of uncertainty described in section 4. Using self-similarity to extend this pattern, it is straightforward to define the permutation operator $i^\alpha$ where α is any number with a finite binary expansion, a so-called dyadic rational. As α gets smaller and smaller, $i^\alpha$ moves elements to the front of the sequence (to the "large scales") from further and further back in the sequence (from the "small scales"). Like the fluid equations in the inviscid limit, $i^\alpha$ has a singular limit as α→0. One way to see this is to imagine that *S* is the binary expansion of a real number, and consider $i^\alpha$ as a function on the reals.

In fact $i^\alpha$ is singular for any α. As a result, $S(\lambda) = i^{2\lambda/\pi}(S)$ is only definable on the circle $0 \leq \lambda \leq 2\pi$ when λ is a dyadic rational multiple of π. More specifically, $S(\lambda)$ cannot be continued to the irrational angles like conventional continuous functions such as $e^{i\lambda}$. We will see the consequence of this shortly.

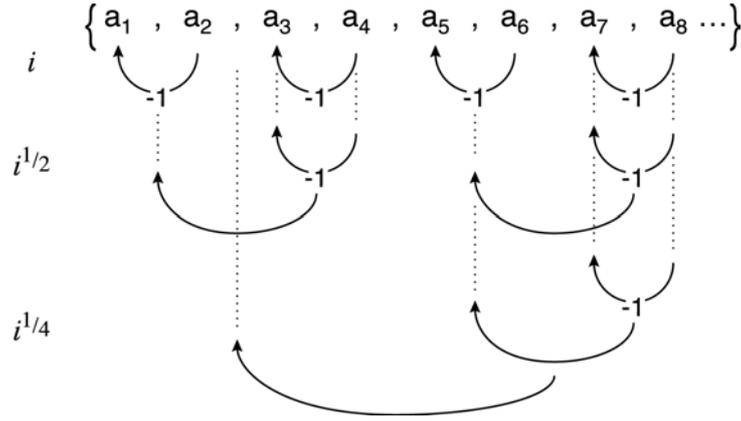

*Figure 4. A schematic of the permutation-operator representation of the unit complex numbers, motivated by the upscale cascade of uncertainty associated with the meteorological butterfly effect.*

Consider now the expression

$$S' = \cos\lambda' \, S + \sin\lambda' \, i(S) \qquad (11)$$

which parallels the familiar additive formula $\cos\lambda + i\sin\lambda$ for a general unit complex number. For simplicity, suppose $0 < \lambda' < \pi/2$. Then we interpret this formula in the present context as follows. If $\cos\lambda'$ is dyadic rational, then

$$S' = \{a'_1, a'_2, a'_3, a'_4, a'_5, a'_6, a'_7, a'_8, ..., a'_{2^N}\}, \qquad (12)$$

in equation (11) denotes a sequence whose correlation with $S$ equals $\cos\lambda'$. Conversely[2], if $\sin\lambda'$ is dyadic rational, then $S'$ denotes a sequence whose correlation with $i(S)$ equals $\sin\lambda'$. As such, $S'$ will not vary continuously with $\lambda'$, and, as before, we can operate on such $S'$ with $i^\alpha$, and represent these on the circle with the formula $S'(\lambda) = i^{2(\lambda-\lambda')/\pi}(S')$.

Consider now the following *key* question: do there exist angles $\lambda$ where both $S(\lambda)$ and $S'(\lambda)$ are simultaneously defined. If $S(\lambda)$ and $S'(\lambda)$ were familiar "classical" functions on the circle (eg $e^{i\lambda}$ and $e^{i(\lambda-\lambda')}$), then both would be simultaneously defined for all angles $\lambda$. It might therefore be naively imagined that by making $N$ in equations (5) and (12) sufficiently large, there would be arbitrarily many angles where both $S(\lambda)$ and $S'(\lambda)$ are simultaneously well defined. However, this is not the case. (The cognoscenti may recognise a similarity between this result and the Heisenberg uncertainty principle in quantum theory, whereby potential measurements based on non-commuting observables cannot simultaneously have well-defined outcomes.) The proof of this result is the central core of this work and makes the sequence construction "non-classical" and plausibly quantum theoretic.

We can prove there are no simultaneously-defined directions by *reductio ad absurdum* using a fundamental number-theoretic property of the cosine function.

---

[2] If $0 < \cos\lambda' < 1$ is dyadic rational, then $\sin\lambda'$ is not, and vice versa.

Assume there are points on the circle at which $S(\lambda)$ and $S'(\lambda)$ are both simultaneously defined; then by the discussion above, $\lambda'$ must be a dyadic rational multiple of $\pi$. For all such $\lambda'$ in the interval $(0, \pi/2)$ it is known that $\cos \lambda'$ is irrational (Jahnel, 2005). However, this contradicts the fact that $\cos \lambda'$ is a correlation coefficient between two finite sequences, and hence must be a rational number. Hence there are no points on the circle at which $S(\lambda)$ and $S'(\lambda)$ are both simultaneously defined.

Using the idealisation of the upscale cascade, as embodied in this reformulation of the complex numbers, we have managed to develop a fundamentally granular (ie generically singular) mathematical structure, without requiring infinite-time integration of the equations of motion to give us the asymptotic notion of a fractal attractor. If quantum theory could be completely reformulated, replacing the standard Hilbert space with sequences of the type $S(\lambda)$, then we might be able to conclude that the evolution of the real universe is not indeterminate, does not have weird superposed states, and is not non-locally causal. Is this possible?

**6. Binary Sequences, Quantum DNA and the Implicate Order**

The discussion so far is leading us to consider reinterpreting the quantum wave-function $|\psi\rangle$ in terms of binary sequences. The individual elements of these sequences would correspond to elements of quantum reality - what John Bell called "be-ables" (Bell, 1993). The proposal is that the sequences define the sample space over which quantum probabilities and quantum correlations are obtained. In such a reinterpretation, the elements of quantum reality never correspond to superpositions of values, but always definite values. Moreover, just as a gene's DNA encodes information about the organism as a whole, $S(\lambda)$ encodes the linkage between $|\psi\rangle$ and environment in which $|\psi\rangle$ belongs.

This notion of an encoded linkage between the parts and the whole, has been referred to by the great 20[th] century quantum theorist David Bohm, as nature's "implicate order" (Bohm, 1980). Bohm's ideas were inspired in part by fluid dynamical phenomena. The notion of implicate order can be grasped by considering equation (2). A partial time series of the *X* component presents an apparently random (or, what Bohm would have called, "explicate") order. On the other hand, with long enough time series and sophisticated nonlinear algorithms, we may be able to reconstruct the encoded implicate order - in this case the underlying fractal attractor.

With this in mind, let us return to the problem of the remote measurements of an ensemble of pairs of entangled spin-1/2 particles as described in sections 2 and 3. The particles themselves belong to a larger system comprising *N* other elementary spin-1/2 systems including, for example, the measuring apparatuses. The implicate order associated with this "belonging" is encoded in $S(\lambda)$.

One of the properties of $S(\lambda)$ is that it is defined on a set of $2^N$ angles, and not on the continuum. However, since in reality N is a very very large number, so that the set of allowed angles appears for all practical purposes to be dense on the circle, such granularity does not, at first sight, appear significant.

However, there is an ontological sense in which this granularity literally makes all the difference in the world. For example, we have seen that the set of angles where $S(\lambda)$ is well defined, is necessarily disjoint from the set of angles where $S'(\lambda)$ is well defined, even though either set is effectively dense on the circle, and even though $S$ and $S'$ may be arbitrarily well correlated with each another. This means that the sample space of directions (eg relative to the distant stars) at which the left hand particles have well-defined states, is effectively dense on the circle, but nevertheless disjoint from the sample space at which the right-hand particle has well defined states (and vice versa). Hence if we ask: what would the state of the left hand particle be relative to one of the directions for which the right-hand particle states are well defined, the answer is "undefined". This is no more than a variant of the argument used in section 3 to rule out counterfactual reasoning (where an arbitrarily-small reorientation of the underlying attractor will take a given point in state space "off the attractor"). As discussed in section 3, if we cannot appeal to counterfactual reasoning, we cannot conclude that quantum phenomena necessarily exhibit non-local causality.

The correlation structure associated with the entangled sequences can also account for quantum interference. For example, from equation (11) the correlation between $S$ and $S'$ varies between -1 (destructive interference) and 1 (constructive interference) as $\lambda'$ is varied.

I want to conclude this section by discussing briefly how this formalism can be extended to describe the wave-function of general multi-state quantum systems. In section 4 we represented the complex number $i$ as a permutation operator acting on successive pairs of elements of a binary sequence $S$. However, this is by no means the only way of representing $\sqrt{(-1)}$ using permutation operators. If instead we consider possible representations of $\sqrt{(-1)}$ based on permutations of successive quadruplets of elements of $S$, then it is easy to find the following set:

$$\begin{aligned} I(S) &= \{-a_2, a_1, a_4, -a_3, -a_6, a_5, a_8, -a_7 ...\} \\ J(S) &= \{-a_3, -a_4, a_1, a_2, -a_7, -a_8, a_5, a_6 ...\} \\ K(S) &= \{-a_4, a_3, -a_2, a_1, -a_8, a_7, -a_6, a_5 ...\} \end{aligned} \quad (15)$$

It can be easily checked that $I^2(S) = J^2(S) = K^2(S) = -S$ and that, in addition, $KJI(S) = -S$. These are the quaternion relations (in operator form). The quaternions were discovered by William Rowan Hamilton, whose name also described an operator in Schrödinger's equation (1). Similar to the construction leading to equation we can also define families of quaternionic sequences eg $S' = \cos\theta\, I(S) + \sin\theta\, J(S)$ which combine elements of the quaternionic sequences $I(S)$ and $J(S)$ such that if $\cos\theta$ is dyadic rational, the correlation between $S'$ and $I(S)$ is equal to $\cos\theta$, whilst if $\sin\theta$ is dyadic rational the correlation between $S'$ and $J(S)$ is equal to $\sin\theta$ (see the footnote above).

Using (15) and the notion of self-similarity, seven further representations of $\sqrt{(-1)}$ can be found from permutations of successive octuplets of $S$. Continuing inductively, the notion of self similarity allows us to generate a set of $2^N - 1$ further $\sqrt{(-1)}$ permutation operators acting on successive $2^N$-tuplets of sequence elements. Briefly, then, a composite system comprising $N$ elemental spin-1/2 systems is represented by $N$ binary sequence $S_j$, $1 \le j \le N$ comprising at least $2^N$ elements; the degrees of freedom being associated with the $2^N-1$ permutation representations of $\sqrt{(-1)}$, which, together with the identity operator, corresponds to that required by quantum theory. The correlation coefficient between pairs of sequences plays the role of the Hilbert space inner product.

The suggestion being made here is that by reinterpreting the conventional continuum field of complex numbers in terms of deterministic granular permutation operators, the quantum wave-function of compound objects (eg you, me, or the cosmos as a whole) can be represented in terms of a set of intricately-encoded binary sequences (somewhat like the DNA of a gene, but in a 4 dimensional space-time setting). Taken one at a time, these sequences look as if they are random (the explicate order) and yet they encode the entanglement of elemental quantum objects as demanded by the conventional Hilbert space representation of quantum theory (the implicate order).

The work described here is essentially a kinematic reformulation of quantum theory. I have not yet discussed how dynamical evolution associated with the Schrödinger equation itself is expressed in this reformulation. This is work for the future. Also, one may legitimately ask whether any of this has any relevance for experimental physics. In this respect it is possible that the present formulation may give a more complete description of entanglement for 4 (or more) entangled spin-1/2 particles, than appears possible in conventional quantum theory. The reason for this relates a theorem in algebraic topology, which states that the Hilbert state space for 4 or more entangled spin-1/2 particles cannot be decomposed ("fibrated" is the correct technical word) into smaller state spaces. The formulation described here, being neither topological nor algebraic, is not restricted by this theorem. Curiously, the graviton (the quantum of the still unknown quantum theory of gravity) can be considered as a composite of 4 spin-1/2 particles.

## 7. Conclusions

Meteorology is an extraordinarily interdisciplinary subject, with quantitative links to many of the applied sciences. However, in this, the 100th anniversary of Einstein's *annus mirabilis*, the possible relevance of nonlinear thinking about predictability of weather has been considered, to try to resolve the two key that led Einstein to ultimately reject quantum theory as a fundamental theory of physics: indeterminacy and non-local causality.

We started by discussing one of the implicit assumptions in the theorem that is conventionally interpreted to imply non-local causality. This implicit assumption is that of counterfactual definiteness. The attractor of the Lorenz (1963) model of low-order chaos was used to illustrate the notion that there may be profound constraints on the freedom to vary local variables in the manner suggested by what might appear intuitively to be sensible (counterfactual) reasoning.

We discussed the self-similar upscale cascade ("meteorological butterfly effect") as a means of overcoming some of the objections to marrying chaos theory and quantum theory. A combinatoric idealisation of the self-similar upscale cascade was then constructed which mimicked complex numbers - the latter playing an essential role in quantum theory's Schrödinger equation. This construction was used to suggest a reformulation of quantum theory where a universe of *N* elemental quantum systems was represented by a set of *N* sequences of binary values. Much as a gene's DNA can be thought of as encoding properties of the organism as a whole, it was suggested that self-similar families of these representations of $\sqrt{(-1)}$ could encode the intricacy of quantum-theoretic entanglement relationships in compound objects.

In conclusion, a case has been made, motivated by nonlinear meteorological thinking, that God does not (or at least need not) play dice, that complex-numbered undead cats do not stalk the earth unobserved, that the universe is not continually splitting into multiple copies, and that the world in which we live, whilst profoundly holistic, does not exhibit non-local causality, or "spooky action at a distance".

## Acknowledgements

The research that led to the work described in this paper has benefited from discussions both with philosophically-minded meteorological colleagues, and with quantum experts from mathematics, philosophy and theoretical physics departments in Europe and the Unites States. Some of these are listed in (Palmer 2004). I am grateful to my meteorological colleagues for numerous helpful discussions. I am also grateful to the real experts for taking the ideas of a professional meteorologist seriously, and most especially for allowing me to present my results in departmental seminars, workshops and conferences. Without this exposure, with all the scrutiny that it implies, I would not have had the confidence to write this paper.

## A Glossary of Terms

**Bell's Theorem**

Based on correlations between measurements on ensembles of pairs of entangled quantum particles. For a large class of putative theories which are locally causal, Bell's theorem requires these correlations to satisfy a certain inequality. The inequality is violated experimentally. As a result, Bell's theorem is usually cited as the fundamental reason Einstein was wrong to believe that non-local causality was not a fundamental feature of quantum physics.

**Complex numbers and quaternions**

The continuum field of complex numbers is widely used in dynamical meteorology. Quaternions are a generalisation of complex numbers and can be related to rotations in 3D physical space. In this paper, both complex numbers and quaternions are represented in terms of self-similar permutation operators acting on the elements of binary sequences.

**Counterfactual Reasoning**

Reasoning about the consequences of something which did not happen, but which intuition suggests might have happened, eg would it have been sunny in London today if a certain butterfly in Amazonia which in reality did flap its wings, had in fact not flapped its wings? An implicit assumption in the derivation of Bell's theorem is that counterfactual questions necessarily have definite answers. We use chaos theory to cast doubt on the validity of counterfactual reasoning.

**Explicate/Implicate Order**

A concept put forward by the 20$^{th}$ century quantum theorist David Bohm, to describe the holistic structure of the physical universe. The implicate order describes some implicit intertwining of the degrees of freedom of a system, not apparent in a partial set of observations. Chaos theory is a manifestation of these ideas - a partial sequence of values of a chaotic variable exhibits random explicit order. With a large-enough sequence and sophisticated algorithms, the encoded implicate order - the underlying fractal attractor - can be revealed

**Finite-time predictability horizon**

An apparent property of 3D fluid equations, at least in the inviscid limit, whereby after some finite time, the large-scale evolution of the system is sensitive to arbitrarily small-scale features in the initial state.

**Indeterminacy/ "Does God Play Dice?"**

One of Einstein's concerns about quantum theory: that during the measurement process, the evolution of the quantum state does not appear to evolve by deterministic laws.

**Low-Order Chaos/ fractal attractor/self-similarity.**

A nonlinear deterministic dynamical system governed by a small number of differential (or difference) equations can give rise to apparent randomness. This apparent randomness is associated with the existence of a fractal attractor in the system's state space. An attractor is a set of points to which state-space trajectories of the underlying dynamical equations are attracted. The fact that the set is fractal means that it has fine scale structure which persists under repeated magnification of the set. Such fractal structure is said to be self-similar. The phenomenon is known as chaos about which volumes has been written.

**Many-worlds interpretation**

Some quantum experts (eg David Deutsch, who developed much of the seminal theory for quantum computing) maintain that each of the possible quantum measurement outcomes (such as the alive /dead cat alternatives in Fig 1) are realised in different parallel universes.

**Meteorological Butterfly Effect**

A description of the loss of predictability in terms of the upscale cascade, rather than low-order chaos. Compared with the volumes written about low-order chaos, relatively little has been written about the meteorological butterfly effect.

**Non-computability**

A class of (mathematically well-posed) propositions is referred to as non-computable if there is no algorithm that can be guaranteed to determine the truth or falsehood of each member of the class. The fractal Mandelbrot set is non-computable. The mathematical physicist Roger Penrose has speculated that a more complete formulation of quantum theory must have non-computable elements.

**Non-local causality/"Spooky Action at a Distance"**

Einstein's principal concern about quantum theory: that a distant measurement could instantaneously cause a change the state of a quantum system "here".

**Schrödinger Equation and Hilbert Space**

The Schrödinger equation is a linear dynamical equation which describes the evolution of the quantum wave-function, at least during periods where the wave-function is not being "measured"; the latter being described conventionally in terms of a random "reduction" process. The state space in which wave-function solutions of the Schrödinger equation reside is known as a Hilbert Space. Because of the existence of $\sqrt{(-1)}$ in the Schrödinger equation, the Hilbert Space is necessarily complex. In quantum theory, the Hilbert Space dimension increases exponentially with the number of quantum particles considered. Any alternate theory must account for not only the complex nature of Hilbert space, but also this exponential "vastness".

**Upscale cascade**

The nonlinear mechanism whereby uncertainty propagates eg in fluid-dynamical equations from small scales to large scales. The mathematical characteristics of the cascade have power-law structure, suggesting scale-invariant (ie self-similar) properties.